\documentclass[aps,prl,reprint,superscriptaddress,nofootinbib]{revtex4-2}

\usepackage{amsmath,amssymb,mathtools}
\usepackage{braket}
\usepackage{graphicx}
\usepackage{hyperref}
\hypersetup{hidelinks}

\newcommand{\Tr}{\operatorname{Tr}}
\newcommand{\id}{\mathbb{I}}

\begin{document}

\title{A Cheat-Sensitive Primitive for Destination-Private Quantum Routing}

\author{S. F. E. Oliviero}
\affiliation{Freie Universit\"at Berlin, Arnimallee 14, 14195 Berlin, Germany}

\author{S. Onofri}
\affiliation{Scuola Normale Superiore, Piazza dei Cavalieri 7, I-56126 Pisa, Italy}

\author{V. Giovannetti}
\affiliation{NEST, Scuola Normale Superiore and Istituto Nanoscienze-CNR, Piazza dei Cavalieri 7, I-56126 Pisa, Italy}

\date{\today}

\begin{abstract}
Quantum networks will rely on intermediate nodes to distribute entanglement, but the same nodes may learn which users are being connected. We introduce a minimal two-call routing primitive in which a user, Alice, establishes an entangled link with one of several clients while testing whether the provider, Bob, has disturbed a coherently encoded request. The construction adapts the hidden-order mechanism of quantum private queries: one call contains a direct request, the other a coherent superposition of the desired and null addresses, and Alice keeps their order secret. A private matching qubit held by the selected client restores the coherent test state after a favorable Bell-measurement outcome, so that Alice can test the returned query without any further operation by Bob. The claim is deliberately limited. The protocol shows that a routing request can carry a physical cheating test, and it detects an explicit destination-reading attack with nonzero probability. It does not yet provide a quantitative information-disturbance tradeoff, composable security, or certification of an unmeasured output against arbitrary attacks. We also characterize the exact perfect-pass limit, in which Bob's residual state, averaged over the private matching states, is independent of the destination.
\end{abstract}

\maketitle

Quantum communication turns physical disturbance into a security signal. Quantum key distribution offers the standard example: an eavesdropper who extracts information from a nonorthogonal encoding necessarily changes observable statistics \cite{BennettBrassard1984,Ekert1991}. In a network, however, protecting the transmitted message is only part of the problem. Intermediate nodes distribute entanglement, join shorter links by entanglement swapping, and choose which endpoint is served \cite{Zukowski1993,Kimble2008,Wehner2018}. In doing so, a router may learn the identity of the intended recipient even when the message itself is perfectly encrypted.
This leads to a basic question: can a user ask an untrusted router for an entangled connection while ensuring that an attempt to read the destination leaves a detectable trace? The problem is related to anonymous quantum communication \cite{ChristandlWehner2005,BroadbentTapp2007}, but the resources are different. Here the intermediate node supplies the bipartite links and performs the routing operation itself; Alice and the selected recipient share no trusted entanglement in advance.

Our starting point is the hidden-order idea behind quantum private queries (QPQ) \cite{Giovannetti2008,Giovannetti2010}. In QPQ, a user hides an ordinary query among nonorthogonal requests and sends them sequentially in a secret order. A provider who tries to distinguish them risks destroying their coherence. Here the query address becomes the destination label, while the database answer is replaced by an entangled link. This change introduces one difficulty: part of the coherent routing request ends at the selected client, so Alice cannot test the request locally. The extra ingredient is a private matching qubit held by that client. After a favorable Bell-measurement outcome, this qubit restores Alice's coherent test state and leaves the same residual state at Bob in the real and null branches. Our aim is to isolate this mechanism in its simplest form. We first describe the honest operation and the projector that implements the cheating test. We then examine a direct address-dephasing attack and show that it is rejected with nonzero probability. Finally, we study the exact perfect-pass regime: if both hidden orders and the requested output are preserved exactly, Bob's residual state, averaged over the private matching states, is independent of the destination. This is a proof-of-principle statement, not a complete security theorem. In particular, the present test neither gives a general quantitative relation between Bob's information and Alice's rejection probability nor certifies the unmeasured output against an arbitrary malicious router.

\begin{figure}[t]
 \centering
 \includegraphics[width=\columnwidth]{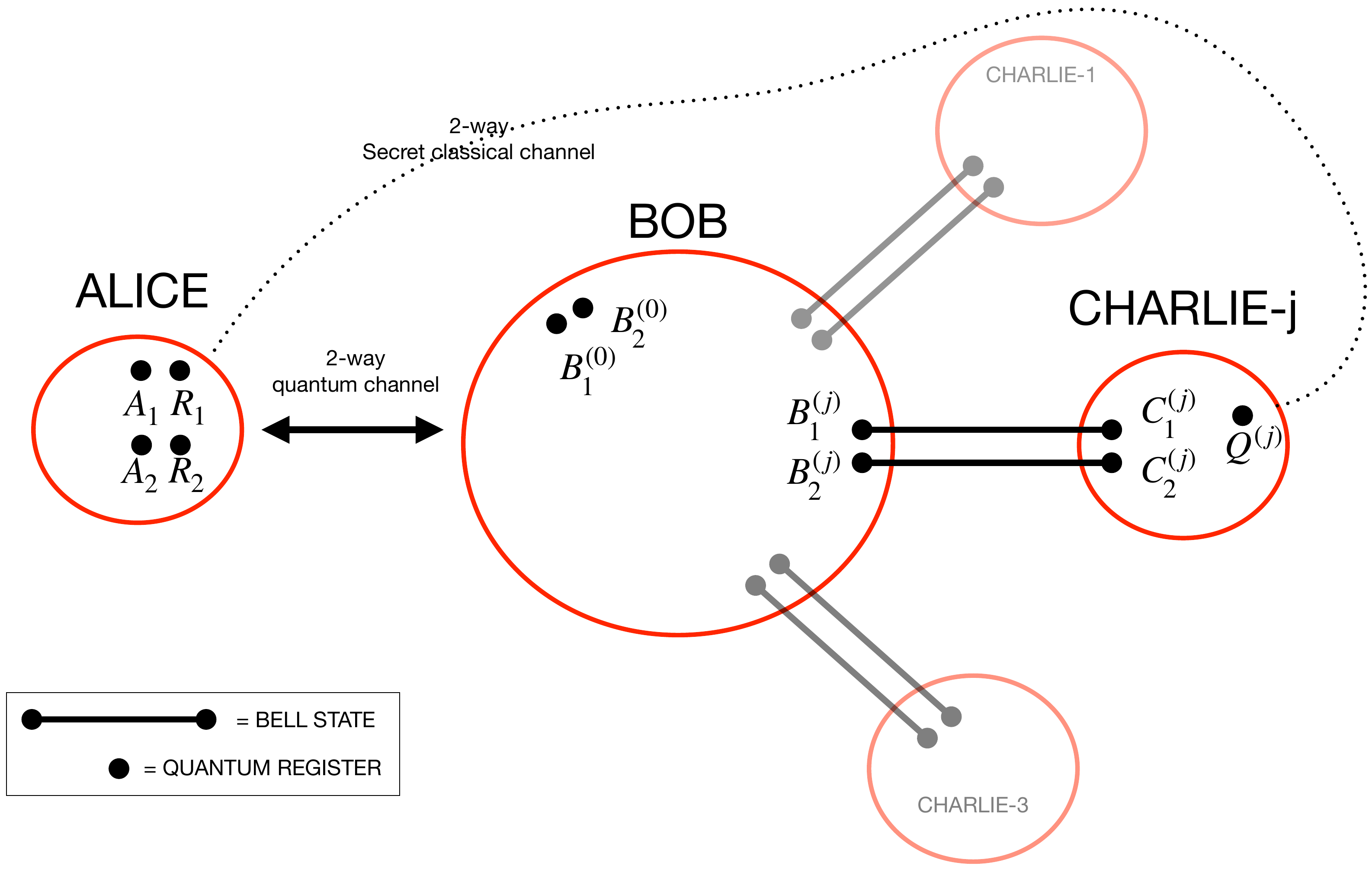}
 \caption{Star-network implementation of the two-call routing primitive. Bob holds two ports $B_1^{(k)},B_2^{(k)}$ entangled with clients $C_1^{(k)},C_2^{(k)}$ at each Charlie-$k$, together with dummy ports $B_1^{(0)},B_2^{(0)}$. The user Alice sends address--payload pairs $(A_s,R_s)$ sequentially. The selected Charlie-$j$ holds a private matching qubit $Q^{(j)}$ and communicates with Alice through a private classical channel. The unselected clients and their links are omitted from the active part of the diagram but remain part of the network.}
 \label{fig:network}
\end{figure}

\textit{Network model and notation.--}\label{sec:model}
The setting consists of a user Alice, a provider Bob, and $N$ clients Charlie-$1,\ldots,$ Charlie-$N$. Alice wishes to establish an entangled qubit pair with a privately chosen Charlie-$j$. To support the request, Bob initially shares two singlets with each client,
\begin{equation}
 \ket{\Psi_-}_{B_s^{(k)}C_s^{(k)}}
 =({\ket{01}_{B_s^{(k)}C_s^{(k)}}
 -\ket{10}_{B_s^{(k)}C_s^{(k)}}})/{\sqrt{2}},
 \label{eq:singlets}
\end{equation}
for $s\in\{1,2\}$ and $k\in\{1,\ldots,N\}$. Bob holds $B_s^{(k)}$, while Charlie-$k$ holds $C_s^{(k)}$. Bob also has two dummy qubits, $B_1^{(0)}$ and $B_2^{(0)}$, both initialized in $\ket{0}$. No Alice-Charlie entanglement is assumed as a trusted resource.
Alice controls two address registers, $A_1,A_2$, and two flying qubits, $R_1,R_2$. Each address register has the orthogonal states $\ket{0},\ket{1},\ldots,\ket{N}$. A nonzero value selects a client, whereas $0$ labels a null request. Alice can exchange each pair $(A_s,R_s)$ with Bob, and one call must be completed before the next begins. In addition, Alice and every Charlie-$k$ share a confidential authenticated classical channel. Its communication pattern is fixed in advance, so the classical traffic does not reveal which client was selected.
Charlie-$j$ also holds a private qubit $Q^{(j)}$. From a shared private classical description, Alice and Charlie-$j$ can prepare the same known state $\ket{\chi}$ in their respective laboratories. Such state is sampled independently from an ensemble satisfying
\begin{equation}
 \mathbb{E}_{\chi}\bigl[\ket{\chi}\!\bra{\chi}\bigr]={\id}/{2}.
 \label{eq:onedesign}
\end{equation}
For this single-qubit construction, one may use the uniform Bloch-sphere ensemble, the computational basis, or the six Pauli eigenstates. The registers and communication pattern are summarized in Fig.~\ref{fig:network}.

For compactness in the following, we write
\begin{align}
 \ket{j\chi}_{AR}&:=\ket{j}_A\ket{\chi}_R,\nonumber\\
 \ket{j\chi+00}_{AR}&:=
({\ket{j}_A\ket{\chi}_R+\ket{0}_A\ket{0}_R})/{\sqrt{2}}.
 \label{eq:querynotation}
\end{align}
The second state coherently combines a genuine routing request with a null request and will serve as the trap.
Furthermore whenever only the selected client is relevant, we abbreviate the initial
router--client resource as
\begin{equation}
 \ket{\Omega_j}_{BCB^{(0)}}=
 \ket{\Psi_-}_{B_1^{(j)}C_1^{(j)}}
 \ket{\Psi_-}_{B_2^{(j)}C_2^{(j)}}
 \ket{00}_{B^{(0)}}.
 \label{eq:omega}
\end{equation}
The unselected ports and their client partners play no role in the
honest derivation and will be left implicit.

\textit{Hidden-order routing protocol.--}\label{sec:protocol}
Alice begins by sampling two independent matching states, $\ket{\chi_1}$ and $\ket{\chi_2}$. She then chooses with equal probability between the following two orders:
\begin{align}
 \text{Type I:}\quad&
 \ket{j\chi_1}_{A_1R_1}
 \ket{j\chi_2+00}_{A_2R_2},
 &Q^{(j)}&=\ket{\chi_2},
 \label{eq:typeI}\\
 \text{Type II:}\quad&
 \ket{j\chi_1+00}_{A_1R_1}
 \ket{j\chi_2}_{A_2R_2},
 &Q^{(j)}&=\ket{\chi_1}.
 \label{eq:typeII}
\end{align}
Including the two singlets, the dummy ports, and Charlie's matching
qubit, the corresponding input states are
\begin{align}
 \ket{\mathrm{I}}={}&
 \ket{j\chi_1}_{A_1R_1}
 \ket{j\chi_2+00}_{A_2R_2}
 \ket{\Omega_j}_{BCB^{(0)}}
 \ket{\chi_2}_{Q^{(j)}},
 \label{eq:fullinputI}\\
 \ket{\mathrm{II}}={}&
 \ket{j\chi_1+00}_{A_1R_1}
 \ket{j\chi_2}_{A_2R_2}
 \ket{\Omega_j}_{BCB^{(0)}}
 \ket{\chi_1}_{Q^{(j)}}.
 \label{eq:fullinputII}
\end{align}
In either case, one call is direct and the other carries the trap, but Bob does not know which comes first. Alice sends $(A_1,R_1)$ and waits until both registers have been returned and isolated before sending $(A_2,R_2)$. This timing is essential: Bob never has access to the two complete queries at the same time.

During call $s$, an honest Bob applies the address-controlled swap
\begin{equation}
 \hat U_{A_sR_sB_s}^{(\mathrm{c\mbox{-}swap})}
 =\sum_{k=0}^{N}\ket{k}\!\bra{k}_{A_s}
 \otimes \hat S_{R_sB_s^{(k)}},
 \label{eq:cswap}
\end{equation}
where $\hat S_{R_sB_s^{(k)}}$ exchanges the states of $R_s$ and $B_s^{(k)}$. Bob then returns $A_sR_s$. On a direct query, Alice's payload is replaced by Bob's half of the selected singlet, producing $\ket{\Psi_-}_{R_sC_s^{(j)}}$, while Bob keeps the matching state $\ket{\chi_s}$ in~$B_s^{(j)}$.
To keep the derivation readable, $\hat U_s^{\mathrm{cs}}$ will denote the full
operator $\hat U_{A_sR_sB_s}^{(\mathrm{c\mbox{-}swap})}$ in
Eq.~\eqref{eq:cswap}.

Once both calls are complete, Alice privately tells Charlie-$j$ which position contains the trap and which matching state should be used. Charlie-$j$ then performs a Bell measurement on $Q^{(j)}C_t^{(j)}$, where $t$ denotes the trap position. We use the Bell basis
\begin{equation}
 \ket{\Psi_{\pm}}=({\ket{01}\pm\ket{10}})/{\sqrt{2}},
 \quad
 \ket{\Phi_{\pm}}=({\ket{00}\pm\ket{11}})/{\sqrt{2}}.
 \label{eq:bellbasis}
\end{equation}
The primitive retains only the favorable outcome $\Psi_-$. Charlie communicates a favorable or unfavorable flag through the private fixed-schedule channel. An unfavorable result is simply inconclusive; it is not, by itself, evidence that Bob cheated. When the favorable result occurs, Alice performs two checks. She verifies that the direct address is $\ket{j}$ and projects the returned trap pair $A_tR_t$ onto $\ket{j\chi_t+00}$. If either test fails, she rejects the run. If both pass, she accepts, and the unmeasured direct pair $R_oC_o^{(j)}$, with $o\ne t$, is kept as the candidate entangled output.

\textit{Honest operation and restoration of the trap.--}\label{sec:honest}
The mechanism behind the test is most easily seen from the teleportation identity
\cite{Bennett1993}
\begin{align}
 \ket{\chi}_Q\ket{\Psi_-}_{XC}
 &=\frac{1}{2}\bigl(
 \ket{\Psi_+}_{QC}\hat Z_X\ket{\chi}_X
 +\ket{\Psi_-}_{QC}\ket{\chi}_X\nonumber\\
 &+i\ket{\Phi_+}_{QC}\hat Y_X\ket{\chi}_X
 -\ket{\Phi_-}_{QC}\hat X_X\ket{\chi}_X
 \bigr), 
\end{align}
which implies 
\begin{eqnarray}
 \bigl(\bra{\Psi_-}_{QC}\otimes\id_X\bigr)
 \ket{\chi}_Q\ket{\Psi_-}_{XC}
 =\ket{\chi}_X/2.
 \label{eq:teleport}
\end{eqnarray}
Let us first follow a Type-I run, where the second call carries the trap.
After the two controlled swaps, the state on the selected route is
\begin{align}
 \hat U_2^{\mathrm{cs}}\hat U_1^{\mathrm{cs}}\ket{\mathrm I}
 ={}&\ket{j}_{A_1}\ket{\Psi_-}_{R_1C_1^{(j)}}
 \ket{\chi_1}_{B_1^{(j)}}\ket{00}_{B^{(0)}}\ket{\chi_2}_{Q^{(j)}}\nonumber\\
 &\otimes\frac{1}{\sqrt2}\Bigl[
 \ket{j}_{A_2}\ket{\Psi_-}_{R_2C_2^{(j)}}
 \ket{\chi_2}_{B_2^{(j)}}\nonumber\\
 &\hspace{13mm}+
 \ket{0}_{A_2}\ket{0}_{R_2}
 \ket{\Psi_-}_{B_2^{(j)}C_2^{(j)}}
 \Bigr].
 \label{eq:honestevolutionI}
\end{align}
Applying Eq.~\eqref{eq:teleport} to the two
terms in square brackets, their component proportional to
$\ket{\Psi_-}_{Q^{(j)}C_2^{(j)}}$ is
\begin{equation}
 \frac{1}{2}\ket{\Psi_-}_{Q^{(j)}C_2^{(j)}}
 \ket{\chi_2}_{B_2^{(j)}}
 \ket{j\chi_2+00}_{A_2R_2},
 \label{eq:favorablecomponentI}
\end{equation}
with the factor $1/2$ being the amplitude of the selected Bell outcome. Therefore, once
that outcome is obtained, the address--payload registers are restored
exactly to the original trap. Including the direct call, the normalized
postselected state hence becomes
\begin{align}
 \ket{\mathrm{I}}\big|_{\Psi_-}
 ={}&\ket{j}_{A_1}\ket{\Psi_-}_{R_1C_1^{(j)}}
 \ket{j\chi_2+00}_{A_2R_2}\nonumber\\
 &\otimes\ket{\chi_1}_{B_1^{(j)}}
 \ket{\chi_2}_{B_2^{(j)}}\ket{00}_{B^{(0)}}.
 \label{eq:honestI}
\end{align}
For Type II the roles of the two calls are exchanged. Before Charlie's
Bell measurement, the state is
\begin{align}
 \hat U_2^{\mathrm{cs}}\hat U_1^{\mathrm{cs}}\ket{\mathrm {II}}
 ={}&\ket{j}_{A_2}\ket{\Psi_-}_{R_2C_2^{(j)}}
 \ket{\chi_2}_{B_2^{(j)}}\ket{00}_{B^{(0)}}\ket{\chi_1}_{Q^{(j)}}\nonumber\\
 &\otimes\frac{1}{\sqrt2}\Bigl[
 \ket{j}_{A_1}\ket{\Psi_-}_{R_1C_1^{(j)}}
 \ket{\chi_1}_{B_1^{(j)}}\nonumber\\
 &\hspace{13mm}+
 \ket{0}_{A_1}\ket{0}_{R_1}
 \ket{\Psi_-}_{B_1^{(j)}C_1^{(j)}}
 \Bigr].
 \label{eq:honestevolutionII}
\end{align}
Projection onto
$\ket{\Psi_-}_{Q^{(j)}C_1^{(j)}}$ again leaves the same
$\ket{\chi_1}_{B_1^{(j)}}$ in the real and null branches. The normalized
state is consequently
\begin{align}
 \ket{\mathrm{II}}\big|_{\Psi_-}
 ={}&\ket{j\chi_1+00}_{A_1R_1}
 \ket{j}_{A_2}\ket{\Psi_-}_{R_2C_2^{(j)}}\nonumber\\
 &\otimes\ket{\chi_1}_{B_1^{(j)}}
 \ket{\chi_2}_{B_2^{(j)}}\ket{00}_{B^{(0)}}.
 \label{eq:honestII}
\end{align}
All four Bell outcomes are equally likely in an honest run. The baseline primitive is therefore conclusive with probability $1/4$. Conditioned on the favorable outcome, both of Alice's checks pass with certainty and the unmeasured direct pair is a singlet.
Notice also that 
Bob is left with $\ket{\chi_1}\ket{\chi_2}$ on the two selected ports: averaging over Alice's independent private choices and using Eq.~\eqref{eq:onedesign} gives
\begin{equation}
 \mathbb{E}_{\chi_1,\chi_2}
 \left[
 \ket{\chi_1\chi_2}\!\bra{\chi_1\chi_2}
 \right]
 ={\id_{B_1^{(j)}B_2^{(j)}}}/{4}.
 \label{eq:mixedports}
\end{equation}
Since the local halves of the unselected singlets are already maximally mixed, once the private matching states are averaged, Bob's complete honest state is the same for every selected port: no information on $j$ is left on Bob side. 

\textit{Cheating test.--}\label{sec:cheating} 
The protocol is cheat sensitive because the direct and trap calls involve nonorthogonal address states. Alice's test is the rank-one projector
 $P_{j,\chi}^{\mathrm{trap}}
 =\ket{j\chi+00}\!\bra{j\chi+00}$.
If, after Charlie's favorable herald, an attack returns the trap state $\rho_{AR}$, Alice rejects with conditional probability
\begin{equation}
 p_{\mathrm{det}}(j,\chi\mid\!\Psi_-)
 =1-\Tr\!\left(P_{j,\chi}^{\mathrm{trap}}\rho_{AR}\right).
 \label{eq:detectiongeneral}
\end{equation}
This is an operational test. Alice may implement the projector by reversing a circuit that prepares $\ket{j\chi+00}$ and then measuring in the computational basis.
To see the test at work, suppose Bob measures the address register of every call in the basis $\{\ket{0},\ket{1},\ldots,\ket{N}\}$ and otherwise follows the honest routing operation. This measurement leaves a direct query unchanged, but it removes the coherence between the real and null branches of the trap. After the favorable Bell outcome, Alice receives
$\rho_{AR}
 =\frac{1}{2}\ket{j\chi}\!\bra{j\chi}
 +\frac{1}{2}\ket{00}\!\bra{00}$ which gives
$ \Tr\!\left(P_{j,\chi}^{\mathrm{trap}}
\rho_{AR}\right)
 ={1}/{2}$.
Thus Alice detects the attack with probability $1/2$, conditioned on Charlie's favorable outcome. Since the favorable outcome itself occurs with probability $1/4$ in the ideal implementation, a single attempt produces both a favorable herald and a detected disturbance with probability $1/8$.
This example establishes the proof-of-principle claim: a routing request can be equipped with a physical test that detects a nontrivial class of destination-reading attacks. A quantitative information-disturbance analysis, together with a certification of the retained output, will be presented elsewhere for a modified, more elaborate version of the scheme designed to exhibit more favorable scaling. Here we pursue a complementary and more basic question: whether the present primitive already enforces destination privacy in the ideal no-disturbance regime.

\textit{Perfect-pass limit.--}\label{sec:perfect}
In the exact perfect-pass limit, the hidden order imposes a strong constraint: Bob's key-averaged residual state is independent of the destination. The result below establishes this statement and makes the role of the hidden order explicit.

Assume ideal initial singlets, sequential calls, and a private hidden order. Let Bob replace the honest swaps by arbitrary quantum operations acting on each received query and his complete memory. Suppose that, for every $j$, every pair $\chi_1,\chi_2$ in the support of the matching ensemble, and both Type I and Type II orders, conditioning on Charlie's favorable outcome yields the exact trap, the correct direct address, and an exact direct singlet. We show that, after averaging over the independent matching states, Bob's complete conditional state is independent of both $j$ and the hidden order. For the present scheme, the exact-direct-singlet assumption is essential to the proof: exact functionality of the direct output is a premise of the proposition, rather than a consequence of the trap test alone. This premise can, however, be relaxed by increasing both the number of singlets shared between Bob and the clients and the number of Alice's quantum queries. For example, with three singlets per client, one trap query, and two direct queries, Alice and the selected client can sacrifice one of the two candidate direct links to test the direct-output functionality of the scheme while retaining the other as the candidate output.

Consider now a dishonest Bob who, instead of applying the controlled-swap operations
$\hat U_{A_sR_sB_s}^{(\mathrm{c\mbox{-}swap})}$, acts on Alice's queries through arbitrary sequential quantum operations. By purifying these operations, we may represent them as unitaries $U_1$ and $U_2$. The first acts on $A_1R_1B$ but not on $A_2R_2$, whereas the second acts on $A_2R_2B$ but not on the already returned registers $A_1R_1$. We define
\begin{equation}
 V_1:=U_1
 \left(\hat U_{A_1R_1B_1}^{(\mathrm{c\mbox{-}swap})}\right)^\dagger .
 \label{eq:defV}
\end{equation}
This change of proof coordinates factors out the ideal first controlled swap from Bob's actual first operation. For compactness, we also write
$X_s:=A_sR_s$ and use
$\ket{\chi_1\chi_2;j}_B$ for the state in which the two selected ports
$B_1^{(j)},B_2^{(j)}$ contain $\ket{\chi_1},\ket{\chi_2}$, respectively,
with the fixed dummy and ancillary registers left implicit.
Charlie's Bell projection commutes with both of Bob's operations, so we
may apply it first.
In Type I it prepares $\ket{\chi_2}_{B_2^{(j)}}$, whereas in Type II it prepares $\ket{\chi_1}_{B_1^{(j)}}$.  
Projecting the remaining direct singlet
onto an arbitrary state on Charlie's side prepares an arbitrary
$\ket{\psi}$ on $R_1$, for Type I, and in $B_2^{(j)}$ for Type II. The perfect-pass
assumption can therefore be written as the following two identities,
valid for all $j,\chi_1,\chi_2,$ and $\psi$:
\begin{eqnarray}
 &&U_2V_1\ket{j\psi}_{X_1}
 \ket{j\chi_2+00}_{X_2}
 \ket{\chi_1\chi_2;j}_{B}\label{eq:constraintI} \\ \nonumber 
 &&\qquad=
 \ket{j\psi}_{X_1}\ket{j\chi_2+00}_{X_2}
 \ket{W_j^{(\mathrm I)}(\chi_1,\chi_2)}_{B}, 
\\
 &&U_2V_1\ket{j\chi_1+00}_{X_1}
 \ket{j\chi_2}_{X_2}
 \ket{\chi_1\psi;j}_{B}\label{eq:constraintII} \\\nonumber
 &&\qquad=
 \ket{j\chi_1+00}_{X_1}\ket{j\psi}_{X_2} \ket{W_j^{(\mathrm {II})}(\chi_1,\chi_2)}_{B}.
\end{eqnarray}
The residual vectors include every system under Bob's control. The
arbitrariness of $\psi$ follows by projecting one half of a singlet onto
an orthogonal qubit state; we absorb the irrelevant phase into its
definition.
Since $U_2$ never acts on $A_1R_1$, the first output in
Eq.~\eqref{eq:constraintI} can be pure for every $\psi$ only if
$V_1$ already factorizes as
\begin{equation}
 V_1\ket{j\psi}_{A_1R_1}\ket{\chi_1\chi_2;j}_{B}
 =
 \ket{j\psi}_{A_1R_1}
 T_j\ket{\chi_1\chi_2;j}_{B}.
 \label{eq:Tj}
\end{equation}
By linearity, $T_j$ cannot depend on $\ket{\psi}$. This is where the
sequential structure enters the proof: once Alice has received the
first query, nothing Bob does during the second call can repair it.

We next expand the Type-II trap in Eq.~\eqref{eq:constraintII}. Its
real branch is governed by Eq.~\eqref{eq:Tj}, while its null branch
initially has an unknown first-call output. Projecting the final
identity onto $\ket{00}_{A_1R_1}$ yields a vector of unit norm. Since
$V_1$ and $U_2$ are isometries, the projection saturates the
Cauchy--Schwarz bound and leaves no component orthogonal to
$\ket{00}_{A_1R_1}$. Hence there is a fixed null-branch map $T_0$
such that
\begin{equation}
 V_1\ket{00}_{A_1R_1}\ket{\chi_1\psi;j}_{B}
 =
 \ket{00}_{A_1R_1}T_0\ket{\chi_1\psi;j}_{B}.
 \label{eq:Tzero}
\end{equation}
Comparing the real and null branches of the same Type-II identity
then gives
\begin{align}
 U_2\ket{j\chi_2}_{A_2R_2}
 T_j\ket{\chi_1\psi;j}_{B}
 &=
 \ket{j\psi}_{A_2R_2}\ket{W_j^{(\mathrm {II})}}_{B},
 \label{eq:realbranchII}\\
 U_2\ket{j\chi_2}_{A_2R_2}
 T_0\ket{\chi_1\psi;j}_{B}
 &=
 \ket{j\psi}_{A_2R_2}\ket{W_j^{(\mathrm {II})}}_{B}.
 \label{eq:nullbranchII}
\end{align}
Because $U_2$ is injective and the product states
$\ket{\chi_1\psi}$ span the selected two-qubit port, it follows that
$T_j=T_0=:T$ on every input subspace used by the protocol:
\begin{equation}
 T_j=T_0=:T.
 \label{eq:Tcommon}
\end{equation}

We can now compare the two calls. Substituting
Eq.~\eqref{eq:Tcommon} into Eqs.~\eqref{eq:constraintI} and
\eqref{eq:constraintII}, and setting $\psi=\chi_2$ in the latter,
gives
\begin{align}
 &U_2\ket{j\chi_2+00}_{X_2}
 T\ket{\chi_1\chi_2;j}_{B}=
 \ket{j\chi_2+00}_{X_2}
 \ket{W_j^{(\mathrm I)}}_{B},
 \label{eq:secondtrap}\\
 &U_2\ket{j\chi_2}_{X_2}
 T\ket{\chi_1\chi_2;j}_{B}=
 \ket{j\chi_2}_{X_2}
 \ket{W_j^{(\mathrm {II})}}_{B}.
 \label{eq:seconddirect}
\end{align}
Expanding the superposition in Eq.~\eqref{eq:secondtrap} and projecting
onto $\ket{00}_{A_2R_2}$ once again saturates the norm bound. Therefore
\begin{equation}
 U_2\ket{00}_{A_2R_2}T\ket{\chi_1\chi_2;j}_{B}
 =
 \ket{00}_{A_2R_2}\ket{W_j^{(\mathrm I)}}_{B}.
 \label{eq:secondnull}
\end{equation}
Equations~\eqref{eq:secondtrap}--\eqref{eq:secondnull} then force
\begin{equation}
 \ket{W_j^{(\mathrm I)}(\chi_1,\chi_2)}_{B}
 =
 \ket{W_j^{(\mathrm {II})}(\chi_1,\chi_2)}_{B}.
 \label{eq:orderindependence}
\end{equation}
Thus, for a fixed destination and a fixed pair of matching states,
Bob's residual state does not depend on the hidden order.

It remains to average over the private matching states. Before
the fixed maps $T$ and $U_2$ act, Bob's averaged state on all labeled
ports is
\begin{equation}
 \overline{\rho}_{B}^{(j)}
 =
 \rho_{\mathrm{anc}}\otimes
 \ket{00}\!\bra{00}_{B^{(0)}}
 \bigotimes_{k\geq 1;s=1,2}
 \frac{\id_{B_s^{(k)}}}{2},
 \label{eq:averagedinput}
\end{equation}
where Eq.~\eqref{eq:onedesign} replaces the two selected matching
states by maximally mixed states; the unselected ports are already
locally maximally mixed. The right-hand side contains no dependence on
$j$. Applying the same fixed quantum operation represented by
Eqs.~\eqref{eq:Tcommon} and \eqref{eq:secondnull} therefore produces
the same averaged residual state for every destination. Together with
Eq.~\eqref{eq:orderindependence}, this proves the proposition. The
argument is exact and supplies no continuity
estimate when Alice accepts only approximately.

\textit{Resources, scope, and limitations.--}\label{sec:scope}
Each attempt uses two Bob--Charlie singlets per client and two sequential Alice--Bob query round trips. Counting both directions, Alice transmits four address qudits and four payload qubits. Charlie-$j$ performs one Bell measurement, and the baseline single-qubit herald is conclusive with probability $1/4$. In an honest conclusive run, the direct position provides one candidate singlet. Every Charlie must also send fixed-length cover messages, ensuring that the classical traffic pattern does not reveal $j$.
The model assumes trusted local devices for Alice and the Charlies, together with private authenticated classical channels. It does not cover malicious clients, loss, repeated use of the same destination, composable anonymity, or implementation noise. More importantly, the trap measurement acts only on the trap position; the direct position remains unmeasured. The protocol therefore demonstrates how a cheating test can be embedded in a routing request, but it does not certify the direct Bell pair against every attack. The perfect-pass proposition applies only when exact direct functionality is required in addition to a passing trap.

These limitations also indicate what a stronger construction would need. One possibility is an additional verification resource; another is a random test/output assignment that prevents the router from knowing, during the protocol, which block will eventually be retained.

\textit{Conclusion.--}
We have distilled a two-query quantum-router protocol into a minimal cheat-sensitive primitive. Alice sends a direct request and a coherent real/null request in a secret sequential order. With the help of a private matching qubit at the selected Charlie, a favorable Bell outcome restores the coherent request in Alice's laboratory, where it can be tested directly. The honest protocol is conclusive with probability $1/4$. For the address-dephasing attack considered here, Alice detects the disturbance with conditional probability $1/2$, corresponding to a joint probability $1/8$ per attempt when the herald is included. In the exact perfect-pass limit, preserving both hidden orders makes Bob's key-averaged residual state independent of the destination.

The present work deliberately stops before a quantitative security threshold. We are preparing an extended version in which the test is supplemented by the resources needed to control the retained output, together with information-disturbance bounds relating the observed rejection probability to the destination information available to a cheating Bob. Under explicit model assumptions, these bounds will allow the observed test statistics to be translated into an estimate of how much information the provider may have acquired.

Language models were only used to edit the
manuscript. All results were derived  by hand by the 
authors.

\end{document}